\documentclass[journal=jpcafh, manuscript=article]{achemso}
\usepackage{amsmath}
\usepackage{graphicx}
\usepackage{subfigure}
\usepackage{multirow}
\usepackage{color}
\usepackage{blkarray}
\SectionNumbersOn

\setkeys{acs}{articletitle = true}

\newcommand{\revision}[1]{{\textcolor{black} {#1}}}

\setkeys{acs}{maxauthors = 10}

\author{Joonho Lee}
\affiliation{Department of Chemistry, Pohang University of Science and Technology, Pohang, Gyeongsangbuk-do, Korea}
\alsoaffiliation{Present Address: College of Chemistry, University of California, Berkeley, California, USA}
\author{Donghyun Lee}
\affiliation{College of Chemistry, University of California, Berkeley, California, USA}
\author{Aleksey A. Kocherzhenko}
\affiliation{College of Chemistry, University of California, Berkeley, California, USA}
\alsoaffiliation{Present Address: Department of Chemistry, California State University, East Bay, USA, USA}
\author{Loren Greenman}
\affiliation{College of Chemistry, University of California, Berkeley, California, USA}
\alsoaffiliation{Present Address: Department of Physics, Kansas State University, 116 Cardwell Hall, 1228 N. 17th St. Manhattan, Kansas, USA}
\author{Daniel T. Finley}
\affiliation{College of Chemistry, University of California, Berkeley, California, USA}
\author{Matthew B. Francis}
\affiliation{College of Chemistry, University of California, Berkeley, California, USA}
\author{K. Birgitta Whaley}
\email{whaley@berkeley.edu}
\affiliation{College of Chemistry, University of California, Berkeley, California, USA}

\title{Molecular Mechanics Simulations and Improved Tight-Binding Hamiltonians for Artificial Light Harvesting Systems: Predicting Geometric Distributions, Disorder, and Spectroscopy of Chromophores in a Protein Environment}

\begin{document}
\begin{abstract}
We present molecular mechanics and spectroscopic calculations on prototype artificial light harvesting systems consisting of chromophores attached to a tobacco mosaic virus (TMV) protein scaffold.
These systems have been synthesized and characterized spectroscopically, but information about the microscopic configurations and geometry of these TMV-templated chromophore assemblies is largely unknown.
We use a Monte Carlo conformational search algorithm to determine the preferred positions and orientations of two chromophores, Coumarin 343 together with its linker, and Oregon Green 488, when these are attached at two different sites (104 and 123) on the TMV protein.
The resulting geometric information shows that the extent of disorder and aggregation properties, and therefore the optical properties of the TMV-templated chromophore assembly, are highly dependent on the choice of chromophores and protein site to which they are bound.
We used the results of the conformational search as geometric parameters together with an improved tight-binding Hamiltonian to simulate the linear absorption spectra and compare with experimental spectral measurements. The ideal dipole approximation to the Hamiltonian is not valid since the distance between chromophores can be very small. We found that using the geometries from the conformational search is necessary to reproduce the features of the experimental spectral peaks.
\end{abstract}

 \maketitle
\newpage
\section{Introduction}

Light harvesting antennae of photosynthetic organisms are exquisitely organized biomolecular structures.\cite{Blankenship2014,VanAmerongen2000}
Although nearly every photosynthetic species on the planet has evolved a light harvesting antenna that is customized to its environment, all these antennae are actually composed of relatively few types of pigment molecules, e.g., chlorophylls, bacteriochlorophylls, carotenoids, phycobiliproteins.
Two additional factors beyond the choice of pigment are critical in the customization of the antennae to very different environments.
These are the tailored structural organization of the pigments, and the tuning of pigment spectral properties by their in-vivo protein environment.
In essence, all LHCs are composed of densely packed pigments that are usually encased in structure-preserving proteins and bound to membranes.
The dense packing of pigments leads to strong electronic coupling between chromophores.
Some LHCs also have a high degree of organization that aligns neighboring dipoles to further enhance electronic coupling.
An example of this is the LH2 system found in purple bacteria, which consists of pigment-protein complexes in which the proteins form helical subunits enclosing rings of 18 and 9 pigments~\cite{Mcdermott1995}.
This strong coupling, along with screening from solvent effects afforded by the binding to photosynthetic membranes, is believed to be the structural basis for the long-lived quantum coherent effects recently observed in a number of light harvesting complexes.\cite{Scholes2017}

Quantum mechanics also plays an important role in the performance of LHCs as antennae for light, both in determining their effective absorption cross-section and the excitation energy transfer subsequent to photon absorption.
As mentioned above, the quantum mechanical coupling {between multiple} pigments can alter the oscillator strength of electronic transitions in pigment-protein complexes.
It is generally believed that this feature helps to increase the efficiency of light absorption in several LHCs.
The most striking example of this comes from green sulfur bacterium, a primitive photosynthetic organism that lives in extremely low light conditions.
Green sulfur bacterium possesses a highly effective antenna structure, the chlorosome, that has recently been identified as large concentric nanotubes of tightly packed bacteriochlorophyll molecules. \cite{Ganapathy2009}
This particular structure leads to very strong inter-pigment coupling and greatly enhanced electronic transition oscillator strengths for efficient light capture and energy transfer.
Much of the drive for construction of artificial light harvesting complexes is to design and construct synthetic molecular complexes that mimic these features of the chlorosome.

The key to producing synthetic mimics of natural light harvesting systems is the establishment of the necessary distance relationships between multiple chromophores.
Although this could, in principle, be achieved using elaborately designed synthetic molecules, this approach is typically quite laborious, is difficult to scale, and leads to highly aromatic systems with poor solubility and limited processing possibilities.
A number of studies have instead used polymers and dendrimers as scaffold materials that establish an upper limit to the distance between chromophores,\cite{Gilat1999,Kim2001} but these systems generally lack the rigidity needed to control transition dipole orientation and to prevent excimer-based quenching pathways.

As an alternative, several groups have developed ways to control the self-assembly of the chromophores themselves, generating large bundles of porphyrins that show energy transfer behavior.\cite{Choi2004,Wang2004}
While these provide interesting chlorosome mimics, it is quite difficult to optimize the performance of these systems to meet specific applications, since the use of new chromophores with different optical properties can lead to unpredictable assembly outcomes.
Alternatively, one can employ self-assembling protein coats of viruses as rigid scaffolds that can template the formation of synthetic light harvesting systems.
In particular, architectures based on the capsid protein monomer of the tobacco mosaic virus (TMV) can be conveniently produced
and employed for the assembly of chromophore arrays by introducing cysteine residues at specific positions
that allow the covalent attachment of a wide variety of commercially-available chromophores with varied optical characteristics.\cite{Miller2007}
One particularly interesting aspect of rod-like light harvesting arrays is the fact that they are inherently three-dimensional, and thus could possess redundant energy transfer pathways that could circumvent defect sites better than linear or ring-like systems.\cite{Miller2010}
An additional advantage of this synthetic system is that the electronic properties of the aggregate complex can be chemically controlled by changing the type of chromophore, the type of linker used to covalently attach the chromophore to the protein,\cite{Delor2018} and the position where it attaches to the protein.


In this work we investigate structural and spectroscopic features of synthetic pigment-protein structures for light harvesting that are based on TMV-templated chromophore assemblies.
The close proximity of the chromophores in the TMV assemblies suggests that their excited electronic states will be closely coupled.
To motivate the design and synthesis of new systems with enhanced electronic coupling, we analyze here several potential synthetic structures using theoretical modeling and spectroscopic characterization.
We employ molecular mechanics simulations of the chromophore-protein systems to provide insight about the geometry and disorder.
This is important given that these are systems for which crystal structures are hard to obtain, and thus direct experimental information about the geometry is lacking.
A key focus of the present study is to understand both the geometry and the mobility of the chromophores, and the extent to which these factors are determined by the microscopic details of the surface of the protein, which typically forces the chromophores to fit into a solvent-accessible pocket.
Different chromophores will be oriented differently and can have varying degrees of mobility depending on their point of linkage and the nature of the link to the protein.
Such geometric and mobility information provides a systematic way to compare and screen for optimal chromophore-protein candidates for synthesis of artificial LHCs.
The geometry of the chromophores is also critical to understanding the optical properties of these aggregate systems, since the electronic coupling between chromophores is primarily determined by the relative orientations of their transition dipole moments (TDMs).\cite{Scholes:2011qf}
In the present work, the geometries of the conformers found from the molecular mechanics simulations are used in a tight-binding model to simulate the optical properties of the system, with comparison to experimental spectra.

The remainder of the paper is constructed as follows.
Section \ref{sec:computation} describes the TMV and chromophore structures employed here and summarizes the computational methods used for the molecular mechanics structural studies with ground state chromophores, as well as the ab initio calculations for electronically excited chromophores and construction of the tight-binding model for simulation of the optical spectra.
Section~\ref{sec:results} presents the structural results with analysis of geometry, orientation and ordering of the chromophores, followed by presentation and analysis of the linear absorption spectra.
Section~\ref{sec:conclusions} concludes with an assessment of the implications for computationally assisted molecular design of artificial light harvesting systems.

\section{Computational Details}\label{sec:computation}
\subsection{The TMV Protein and Chromophores}



\begin{figure}
    \begin{center}
        \subfigure[~Coumarin 343 + Ethyl-Maleimide]{
        \includegraphics[scale=1]{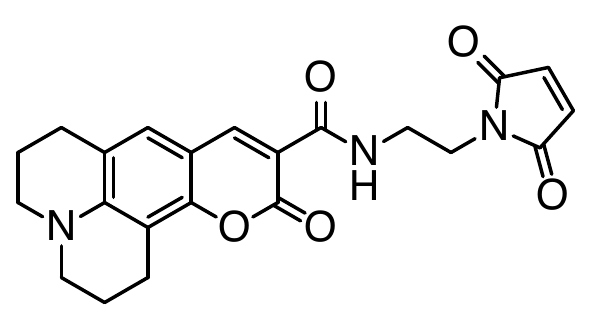}} \qquad
        \subfigure[~Oregon Green 488]{
        \includegraphics[scale=1]{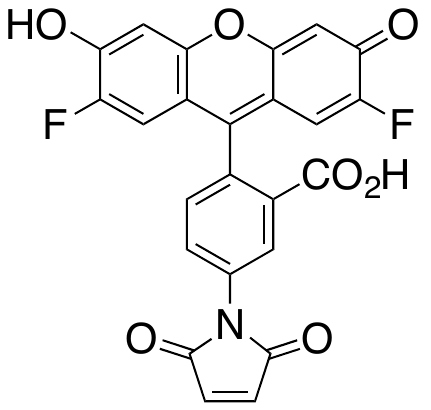}}
    \end{center}
    \caption{\small
    Molecular structures of modified Coumarin 343 (i.e., Coumarin 343 with a linker molecule, ethyl-maleimide) and Oregon Green 488.
    \label{fig:chromophores}}
\end{figure}

The chromophore-protein complexes studied in this paper have all been experimentally synthesized. \cite{Finley2014}
We have successfully attached chromophores to the TMV protein and the theoretical study of {these complexes} is the focus of this paper.
The details of the self-assembly of TMV are available in ref. \citenum{Miller2007} and the details of attaching chromophores to the TMV protein are presented in ref.~\citenum{Finley2014}.

\begin{figure}
    \begin{center}
        \subfigure[~104 position]{
        \includegraphics[width=220px]{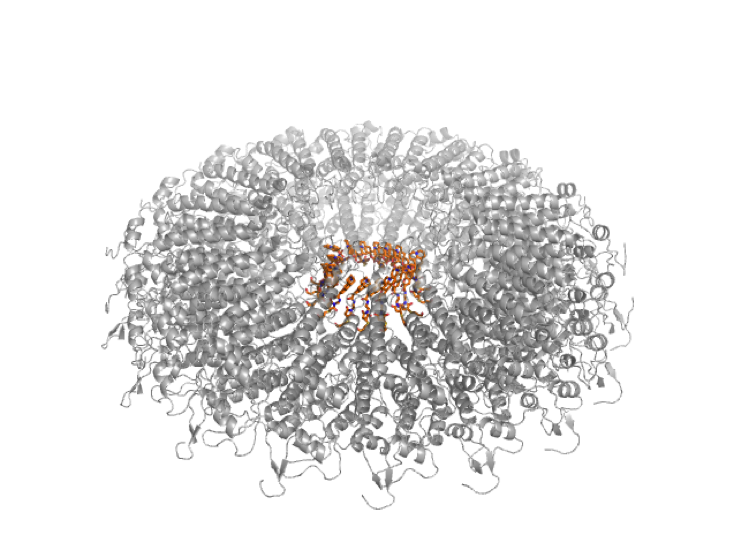}} \qquad \subfigure[~123 position]{
        \includegraphics[width=180px]{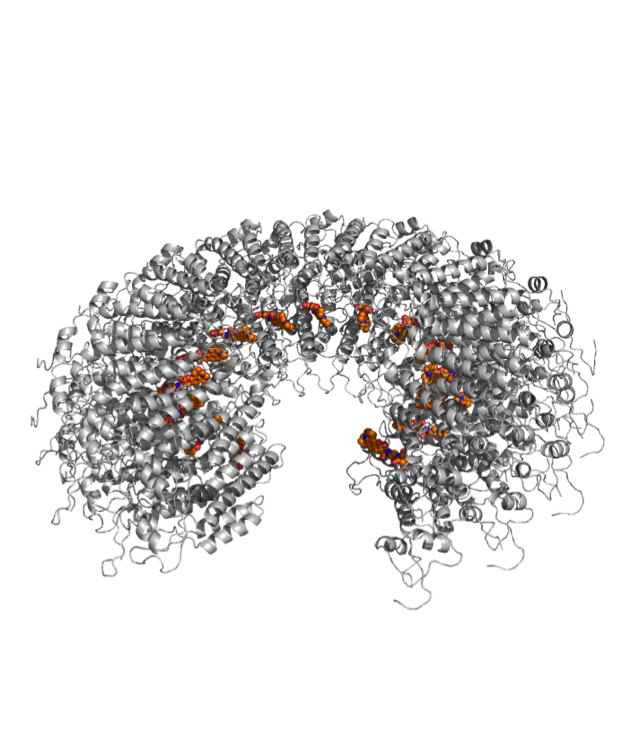}}
    \end{center}
    \caption{\small
    TMV-chromophore double-disk system. Protein colored grey, and chromophore colored orange.
    (a) Chromophore attached at the 104-residue position (inner ring)
    (b) Chromophore attached at the 123-residue position (outer ring)
    \label{fig:residue_positions}}
\end{figure}

The TMV systems were self-assembled into a double-disk with 17-fold radial symmetry. 
We {have} studied the chromophores, Coumarin 343 and Oregon Green 488 (OG), {attached to} the TMV protein at either the 104 (inner ring) or 123 (outer ring) residue positions.
OG can be attached directly to the residues without modification while Coumarin 343 requires a linker molecule.
Coumarin 343 was attached to both of the 104 and 123 positions with an
ethyl-maleimide moiety.
We refer to this complex of Coumarin 343 {together with a linker molecule} as CE.
The molecular structures of CE and OG are available in Figure~\ref{fig:chromophores}.
The 104 and 123 positions differ in {their} distance from the center of the disk, thereby controlling the distance between neighboring chromophores, as illustrated in Figure~\ref{fig:residue_positions}.
These systems will henceforth be referred to as CE-104, CE-123, OG-104, and OG-123.

\subsection{Conformational search using Monte Carlo Multiple Minimum}
The TMV-chromophore system is rather complex and {is impossible to treat fully quantum-mechanically with today's computational resources.}
Therefore, we {explore the high-dimensional configurational} space of the system using the Monte Carlo Multiple Minimum (MCMM) algorithm with the force field of OPLS2005~\cite{Kaminski2001}.
All molecular mechanics simulations presented here were run with the Schrodinger's MacroModel software suite.\cite{MacroModel}
The double disk system has 34 monomers arranged with 17 monomers per layer, which includes {approximately} 200 rotatable bonds.
For the simulations presented {below}, we focus on {a single} layer.
An MC search over the full parameter space is computationally intractable, {since} the required time scales exponentially with the number of rotatable bonds.

{For the CE-104 system we carried out computationally intensive simulations with the full 17 monomer ring system and compared this with simulations in which we considered only five monomers on each layer in the TMV 
under the assumption that chromophores that are separated by two or more monomers are non-interacting. We found that the  average parameters (position and orientation) do not change significantly by using the truncated system. We therefore simulated only the truncated system for the other systems, CE-123, OG-104, and OG-123, and made the spectroscopic analysis on all four systems from the MCMM configurations, using only the middle three chromophores on the top layer for the subsequent structural analyses}.

{The active space of the configurational search, are the atoms constituting the chromophore and linker molecules: these are represented by freely movable atoms.} 
All {protein} atoms within 14 residues {away from} the chromophore constitute the first layer of nearby atoms.  These are not freely movable, but are constrained to their initial position using a harmonic potential with a force constant of 200 kJ mol$^{-1} \text{\AA}^{-2}$.  All protein atoms between 15 and 29 residues away from the chromophore constitute the second layer of nearby atoms: these are effectively frozen in place using a stronger harmonic constraining potential.
All protein atoms further away were ignored, as they are too distant to have any significant interaction with the free atoms.
{In the Monte Carlo molecular mechanics algorithm, given a starting molecule configuration, one randomly chooses some rotatable bonds (these are identified on the basis of the above specifications) and then rotates the atoms about these by a random angle.
The potential energy of the resulting configuration is evaluated according to the force field, and the configuration accepted if the value lies within an acceptable energy window. This yields new conformations of the molecule. The geometry of the new structure is then minimized to find a local minimum, using the Polak-Ribiere Conjugate Gradient (PRCG) method \cite{Polak1969}.} 
{For each configuration, we identify the coordinates of the chromophores and calculate the center of mass and the orientation of the TDM.
To approximate the transition dipole moment for each chromophore, we pick three atoms on the perimeter of the conjugated rings of the chromophores to define a molecular plane.
The normal vector to this plane is calculated for both the ground state DFT-optimized geometry and the MCMM geometry, and the rotation matrix that connects these two normal vectors is then applied to the TDM vector obtained from TDDFT (see Table \ref{tbl:coumarin343}).
Since the conjugated rings generated by the MCMM simulations are quite rigid, 
validating the mapping of the MCMM configurations onto the DFT-optimized molecular structures,
this is a cost-effective way to approximate the TDDFT TDM values of chromophores in the configurations generated by the MCMM calculations.}

\subsection{Ab-initio calculations of excited states}
For the spectral simulations shown later, we {need} transition dipole moments (TDMs). This was achieved by performing time-dependent density functional theory (TDDFT) \cite{Dreuw2005} with B3LYP\cite{Becke1993}/6-31G(d) \cite{Petersson1988,Petersson1991} {for the DFT-optimized molecular geometries}. {Adiabatic excitation energies were extracted as the difference of TDDFT and DFT calculations at the DFT-optimized geometries.}
We truncated the linker molecule of CE and replaced the {ethly-maleimide} linker with a methyl group for simplicity. 
{The} TDDFT calculations employed 75 radial grid points and 302 Lebedev angular grid points.
We also employed equation-of-motion coupled-cluster singles and doubles (EOM-CCSD) \cite{Krylov2008} to further verify  excitation energies and TDMs within the same basis set.
These calculations were run with the development version of Q-Chem. \cite{Shao2015}

\begin{table}[htp!]
\caption{\small
{Electronic structure results for the Coumarin-343 molecule with a methyl group linker substitute, showing
excitation energies and corresponding ground-excited state transition dipoles and oscillator strengths, obtained from TDDFT-B3LYP, and EOM-CCSD calculations.}
\label{tbl:coumarin343}}
\centering
\begin{tabular}{ll|r|r}
                             &   & TDDFT-B3LYP & EOM-CCSD  \\
\hline
first excitation energy (eV) &   &  3.4742     &  3.7753\\
\hline
transition dipole (a.u.)   & x &  2.6354     &     2.7087     \\
                             & y &    0.0057   &           0.1280\\
                             & z &  0.0060     &          -0.0231 \\
\hline
oscillator strength (a.u.)          &   &  0.5912     & 0.7609
\end{tabular}
\end{table}

{In Table~\ref{tbl:coumarin343}, we present the first excitation energy of the Coumarin-343 molecule with a methyl group linker substitute, {obtained from} the TDDFT-B3LYP and EOM-CCSD methods. We do not present the absolute energies of individual electronic states, since for TDDFT-B3LYP only the relative energies are meaningful, while {the} EOM-CCSD calculations are not converged with respect to the basis set size.
In addition to the similar excitation energies and TDMs obtained from TDDFT-B3LYP and EOM-CCSD, the two methods also yield similar wavefunctions in terms of dominant electronic configurations.
Since the first excited states in both chromophores are of singly-excited open-shell singlet character, the excitation energies from both of these two electronic structure methods are expected to be very accurate.
The largest source of error is likely the limited size of basis set employed here, but in a previous paper \cite{Lee2013} it was
shown that employing a larger basis such as 6-311G* does not significantly affect the excitation energies (the change is 0.06 eV for both EOM-CCSD and TDDFT-B3LYP).}

{For the spectral simulations, we used the TDDFT-B3LYP energies and TDMs, rather than the corresponding EOM-CCSD values, for the following reasons.
First, the TDDFT-B3LYP calculation is expected to be closer to its complete basis set limit, since the EOM-CCSD calculation is more sensitive to increasing the basis set size.
Second, it was pointed by Koch et al. \cite{koch1994} that EOM-CCSD does not yield size-intensive TDMs, and thus EOM-CCSD may become a less reliable way to obtain TDMs for large systems.
For these reasons, the analysis requiring TDMs (i.e., the spectral calculations) were all carried out using TDDFT-B3LYP rather than EOM-CCSD.}

\subsection{Spectral Simulations}\label{ssec:spectra}
\subsubsection{Hamiltonian Parametrization}
A tight-binding Hamiltonian of the chromophores is often used to describe the electronic and optical properties of chromophore-protein systems\cite{Ame.Val.etal-2000, Scholes2006a} and we {employ} a similar model {here}:

\begin{equation}
\label{eqn:hamiltonian}
\mathrm{\hat{H}} = \sum_{i=1}^N\varepsilon_i \mathrm{\hat{a}}^\dagger_i \mathrm{\hat{a}}_i +  \sum_{i = 1}^N \sum_{j=1}^{N} J_{ij} \mathrm{\hat{a}}^\dagger_i \mathrm{\hat{a}}_j,
\end{equation}
where $\epsilon_i$ is the on-site energy and $J_{ij}$ is the coupling parameter between the site $i$ and the site $j$. The coupling $J_{ij}$ is a function of the positions of chromophores $i$ and $j$, and {of} the relative orientations of each of their transition dipole moments.
The close proximity between some of the chromophores in our system means that we cannot expect that the commonly used ideal dipole approximation (IDA) \cite{Kasha1965} for $J_{ij}$ couplings {will hold for all configurations}.

We define $\Delta E_n^\text{dimer}$ to be the $n$-th electronic excitation energy of the dimer and $\Delta E_n^\text{monomer}$ to be the $n$-th electronic excitation energy of the monomer.
In the case of a well-separated dimer, both $\Delta E_1^\text{dimer}$ and $\Delta E_2^\text{dimer}$ approach to $\Delta E_1^\text{monomer}$ and therefore the average of  $\Delta E_1^\text{dimer}$ and  $\Delta E_2^\text{dimer}$ is identical to  $\Delta E_1^\text{monomer}$.
However, our previous work shows that at close distances ($<$ 12 \AA), the average will begin to deviate from the monomer excitation energy ($\Delta E_1^\text{monomer}$).
The extent of the deviation will depend on the distance and relative orientations of the two interacting chromophores \cite{Lee2013}.

In order to account for {this} geometry-dependent effect, we obtain more accurate Hamiltonian parameters based on pairwise TDDFT, as follows.
{The off-diagonal elements are given by}
\begin{equation}
\label{eqn:ham_param_j}
J_{ij} = J^\text{TDDFT}(\vec{R}_i,\vec{\mu}_i,\vec{R}_j,\vec{\mu}_j) = \frac{\Delta E_2^\text{dimer} - \Delta E_1^\text{dimer} }{2},
\end{equation}
and the diagonal elements are
\begin{equation} \label{eqn:ham_param_e}
\varepsilon_i = \Delta E_1^\text{monomer} + \Delta V_i
\end{equation}
where
\begin{equation}
\Delta V_i = \sum_{j\ne i} V^\text{TDDFT}(\vec{R}_i,\vec{\mu}_i,\vec{R}_j,\vec{\mu}_j),
\end{equation}
and
\begin{equation}
V^\text{TDDFT}(\vec{R}_i,\vec{\mu}_i,\vec{R}_j,\vec{\mu}_j) = \frac{\Delta E_2^\text{dimer} + \Delta E_1^\text{dimer} }{2} - \Delta E_1^\text{monomer},
\end{equation}
{with $\vec{R}_i$ the position of the $i$-th chromophore and $\vec{\mu}_i$ the orientation} of the $i$-th chromophore.

{In the full data set produced by the} molecular mechanics configurations, there are about $10^7$ pairwise interactions.
Running a TDDFT calculation for each pairwise interaction in every configuration instance is computationally intractable and also redundant, since many of the pairs will have similar geometries.
Our approach is therefore to instead run a TDDFT calculation at selected distances and orientations, and {to} interpolate between the values from these calculations to predict ab-initio values for other geometries.
The precise procedure is as follows:
\begin{enumerate}
\item We parametrize the relative orientation of two interacting monomers: $r, \theta_A, \theta_B, \phi_B$ as defined in ref. \citenum{Lee2013}.
\item We then discretize the space along those variables and calculate the TDDFT energies at the geometries defined by the following grid points:
    r = [5 \AA, 5.25 \AA, 5.75 \AA, 6 \AA, 6.25 \AA, 6.5 \AA, 6.75 \AA, 7 \AA, 7.5 \AA, 8 \AA, 8.5 \AA, 9 \AA, 10 \AA, 12 \AA, 14 \AA],
    $\theta_A$ = [-90$^\circ$, 90$^\circ$] with a 15$^\circ$ increment, $\theta_B$ = [0$^\circ$, 180$^\circ$] with a 15$^\circ$ increment, and $\phi_B $= [0, 180] with 30$^\circ$ increment.
    Out of the possible 18928 geometries, we discard the points that yield unphysical geometries, which {results} in a training set of 4456 energies.
\item We employ a model function ({\it vide infra}) with three free parameters each for $J_{ij}$ and $V_{ij}$. We fit these parameters by using linear regression to match $J^\text{TDDFT}$ and $V^\text{TDDFT}$, respectively, at each of the grid points.
\end{enumerate}

The model functions used to describe $V$ and $J$ are
\begin{align}
\label{eqn:interpolated_coupling}
J_{ij}^\text{model}(r, \theta_i, \theta_j, \phi_j) &= C_J(r)J_{ij}^\text{IDA}(r, \theta_i, \theta_j, \phi_j)\\
V_{ij}^\text{model}(r, \theta_i, \theta_j, \phi_j) &= C_V(r)V_{ij}^\text{IDA}(r, \theta_i, \theta_j, \phi_j)\\
C_k(r) &= \frac{c_1^k}{ (c_2^k-\exp(r/c_3^k))}{, ^k = J,V}
\end{align}
where $J_{ij}^{IDA}$ and $V_{ij}^{IDA}$ are the couplings obtained from IDA, and $C_k$ is a {three-parameter logistical function whose value indicates whether the IDA brakes down. Specifically,}
if $C_J$  and $C_V$ deviate significantly from 1, then that is precisely when IDA breaks down.
The fitted parameters are found to be:
$c_1^V = 0.10388804$, $c_2^V=0.21424357$, $c_3^V=2.2014798$
$c_1^J = 0.2889955$, $c_2^J=1.50871422$, $c_3^J=3.14407206$.

Table~\ref{tbl2:ham_avg} shows the average values of $\Delta V_i$ and $J_{ij}$ obtained using this procedure for Coumarin 345 attached to TMV disks by the methyl group linker in an MCMM sampling of 10$^5$ molecular configurations.

\begin{table}[ht]
\centering
\begin{tabular}{|c|r|r|r|r|r}
\hline
            & $\langle \Delta V_i \rangle$ [eV] & $\langle J_{ij} \rangle$ [eV] 
            \\ \hline
CE-104 & 0.042 & $2.00\times 10^{-3}$ 
\\ \hline
CE-123 & 0.006 & $3.44\times 10^{-4}$ 
\\ \hline
\end{tabular}
\caption{Effective Hamiltonian parameters for {the Coumarin 345 and methyl linker system}, averaged over all sites and 5000 MCMM geometry configurations on TMV double disks. $V_{ij}$ are the diagonal Hamiltonian matrix elements (Eq. \eqref{eqn:ham_param_e}), and $J_{ij}$ are the off-diagonal Hamiltonian matrix elements (Eq. \eqref{eqn:ham_param_j}).}.
\label{tbl2:ham_avg}
\end{table}

\subsubsection{Linear Absorption Spectra}
In order to simulate the linear absorbance of the full double disk TMV system, we first sample small slices (i.e., 5 monomers) from the MCMM configurations and concatenate the middle three chromophores to generate the full system.
Since 17 is a prime number, we need to take 5 samples of three chromophores, and 1 sample of two chromophores, all of which are sampled randomly.
This is repeated for both the upper and lower disks.
For each geometry sample, we extract the center of mass positions and the transition dipole moments ($R$ and $\mu$) of each chromophore.
Next, the tight binding Hamiltonian in Eq.~\eqref{eqn:hamiltonian} is constructed using the parameters $\varepsilon_i$ and $J_{ij}$ described in Eq. \eqref{eqn:interpolated_coupling}.
The Hamiltonian is diagonalized {by a unitary transformation} to yield exciton states and energies:
\begin{align}
    \hat{H}|\psi_k\rangle &= E_k|\psi_k\rangle\\
    |\psi_k\rangle &= \sum_i^N c_{ik}|\phi_i\rangle.
\end{align}

The linear absorption spectrum for a given Hamiltonian is then calculated using
\begin{align}
    \label{eq3:exciton_tdm}
    \vec{\mu}_k &= \sum_i^N c_{ik} \vec{\mu}_i\\
    \label{eq:spectra}
    \text{Abs}(E) \propto \sum_k^N &\left\lVert\vec{\mu}_k\right\rVert^2 \exp\left[-\frac{(E- E_k)^2}{2\sigma^2}\right],
\end{align}
where $\vec{\mu}_k$ is the transition dipole moment of exciton $k$, {obtained by transforming the vector of molecular transition dipole moments  with the same unitary transformation that diagonalizes the Hamiltonian.}
The summation in Eq.\eqref{eq:spectra} describes a discrete convolution between a gaussian function, and a stick spectrum composed of excitations from the ground to single excitation states of the Hamiltonian, with weights given by the 2-norm squared of the exciton transition dipole moment.
The variance of the gaussian function ($\sigma$) is the line broadening parameter for our simulated linear absorption spectrum at a given geometry configuration and corresponds to the homogeneous linewidth.
In the limit where $\sigma\to0$, we recover the eigenvalue stick spectrum.
Eq. \eqref{eq:spectra} yields the spectrum for a single geometry instance.
This process is repeated 5000 times (large enough to obtain converged spectra) to average over the different possible geometry configurations, and is then normalized by the maximum absorbance.


\section{Results and Discussion}
\label{sec:results}
\subsection{Geometric Distributions of Chromophores}
We analyze the MCMM conformations based on the center of mass (CM) positions of the chromophores and the orientations of TDMs of the chromophores.
Those two collective variables are particularly useful in understanding the geometric distribution, as we shall see below.

\begin{figure}[ht]
        \includegraphics{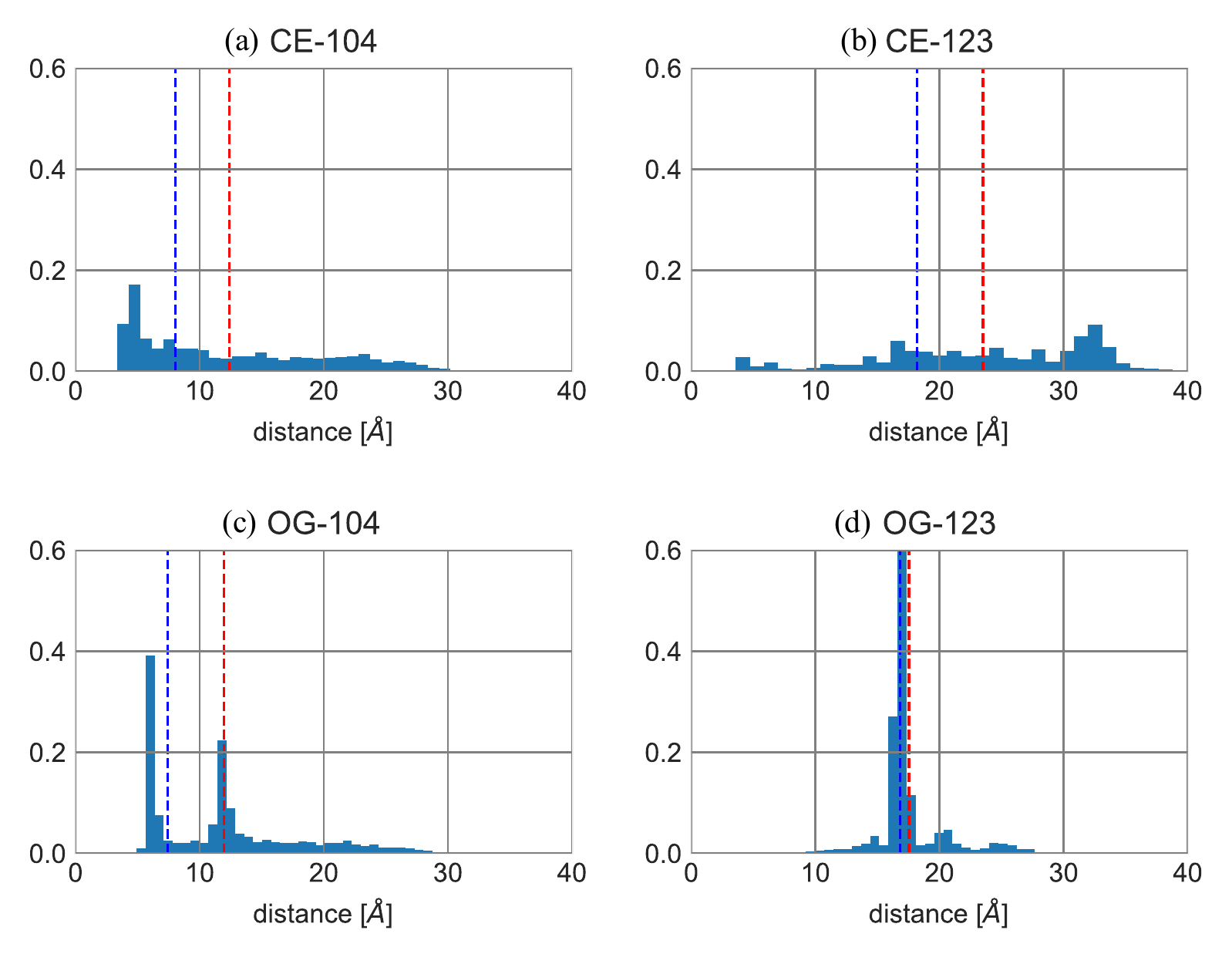}
    \caption{\small
    Normalized histograms of the center-of-mass distances between nearest-neighboring chromophores in
    (a) CE-104, (b) CE-123, (c) OG-104, and (d) OG-123.
    The blue dotted line {in each panel indicates the ``ideal" nearest-neighbor distance, $r_\text{ideal}$, and the red dotted line indicates the mean distance, $\bar{r}$ (see text).}
    The number of samples used in each histogram is 3607, 5499, 5918, and 5810, respectively.
    \label{fig:nn_hist}}
\end{figure}

Figure \ref{fig:nn_hist} shows a histogram of distances between the CM positions of nearest-neighboring chromophores.
We first note that both CE-104 and CE-123 exhibit significant multimodal behaviors, while bimodal and monomodal behaviors are observed for OG-104 and OG-123, respectively.
The qualitative difference between CE and OG can be explained simply: the linker molecule in CE allows Coumarin to move easily its CM position whereas OG has no linker molecule in our study.
We further computed the ``ideal'' nearest-neighbor distance, $r_\text{ideal}$, which assumes an equilateral 17-polygon, and the mean of nearest-neighbor distances, $\bar{r}$.
Those two values are not meaningful in the case of highly multimodal histograms as in the CE cases.
In the case of OG-123, two values are almost identical, whereas each of two peaks in OG-104 roughly corresponds to $r_\text{ideal}$ and $\bar{r}$.

The significance of Figure \ref{fig:nn_hist} is that some chromophores (in particular CE-104 and OG-104) in the TMV systems are not far enough {apart for the IDA to be valid}; the distance between nearest neighbors is often less than 12 \AA.
Based on our previous study of Coumarin 343, when two chromophores are closer than 12 \AA, it is likely that IDA to the Hamiltonian starts to fail quite catastrophically.\cite{Lee2013}
This was indeed our motivation to go beyond IDA, and this will be discussed further {below}.
We note that in the previous study{,\cite{Lee2013} Coumarin was considered without a linker molecule.}
However, we expect the failure of dipole approximations to appear similarly {for Coumarin attached to a linker molecule}.
We {can also expect the same behavior for the OG chromophore}.

\begin{figure}[ht]
    \begin{center}
        \includegraphics[scale=1.0]{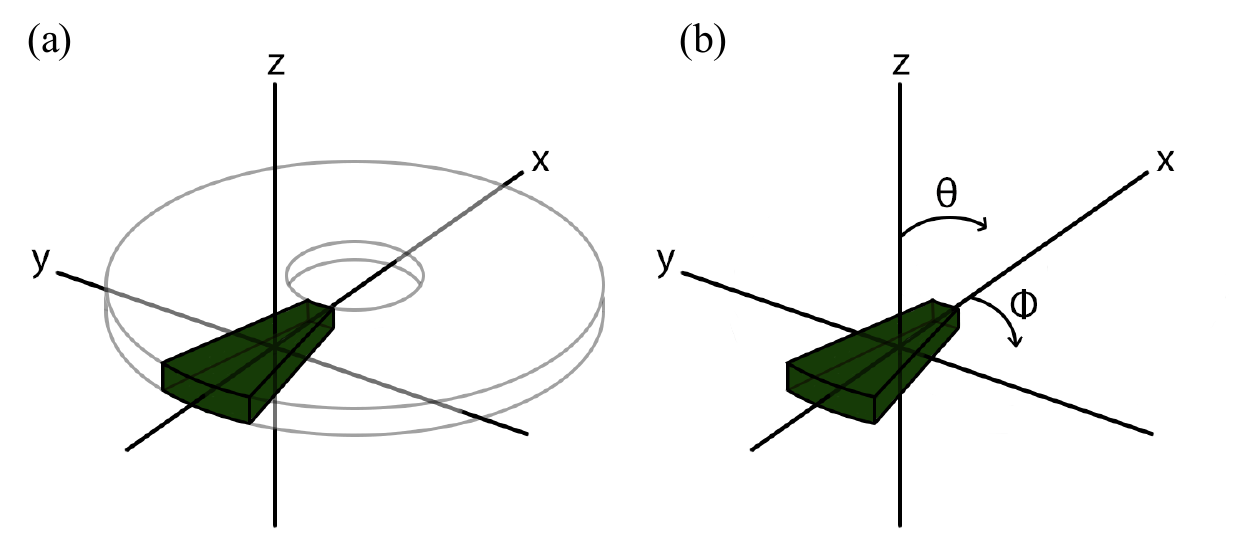}
    \end{center}
    \caption{\small
    (a) Schematic description of the monomer frame embedded into the entire TMV represented by a disk. The monomer is represented by the wedge.
    (b) Schematic description that represents the polar angle $\theta$ and the azimuthal angle $\phi$ in the monomer frame.
    \label{fig:monomer_frame}}
\end{figure}
For the purpose of analysis, we introduce the monomer frame illustrated in Figure~\ref{fig:monomer_frame}.
We define the monomer frame as follows: for every chromophore, the x-axis points towards the center of TMV disk, the z-axis is parallel to the axis of rotational symmetry, and the y-axis is defined in the conventional way {given these x- and z-axes} (i.e., $\hat{y} = \hat{z}\times\hat{x}$).
The radial axis of the polar plot ranges from 0$^\circ$ to 180$^\circ$ and corresponds to the polar angle of the monomer frame.
The angular axis of the polar plot ranges from 0$^\circ$ to 360$^\circ$ and corresponds to the azimuthal angle of the monomer frame.

\begin{figure}[ht]
    \begin{center}
        \includegraphics[scale=0.9]{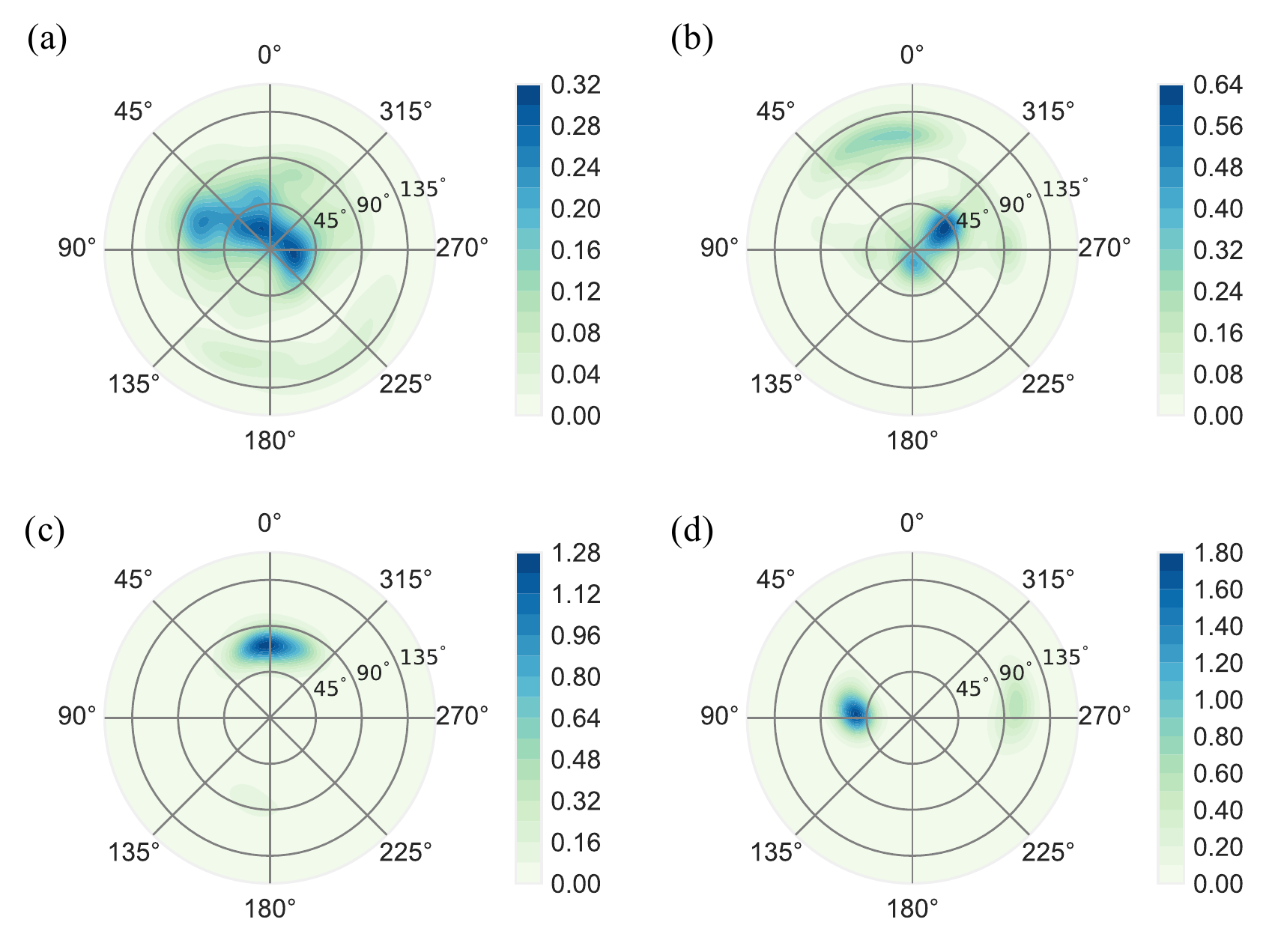}
    \end{center}
    \caption{\small
    Normalized distributions of the transition dipole moments orientation for (a) CE-104, (b) CE-123, (c) OG-104, and (d) OG-123.
    The radial distance indicates $\theta$ and the angular orientation indicates $\phi$. $\theta$ and $\phi$ are defined in Figure \ref{fig:monomer_frame}.
We fitted the TDM vectors to a bivariate gaussian kernel density estimator with a 0.2 bandwidth to obtain this figure.
    \label{fig:tdm_kde}}
\end{figure}

Figure \ref{fig:tdm_kde} shows a histogram of the orientation of TDMs measured in the monomer frame.
It shows that the {TDMs for the CE systems have broader angular distributions than the TDMS in the OG systems}.
In other words, OG systems are far more confined {in their orientation than CE systems}.
This does not necessarily mean that OG systems are more ordered. {The quantification of order-disorder will be discussed later in the paper.
Instead, the wider distributionof the TDMs in the CE systems can be understood by recognizing} the effect of the presence of a linker molecule.
We see similar trends in both chromophores when attached to the 104 or 123 position.
The 104 position exhibits a wider vertical spread compared to the 123 position, whereas the 123 position shows a wider horizontal spread compared to the 104 one.
The broad distribution of the CE systems is somewhat surprising, given that the chromophore molecules are surrounded by the TMV protein environment.
{However, the linker molecule gives enough flexibility to the Coumarin chromophore which results in a broad distribution of geometries}.
This is one of the reasons that make atomistic simulations of the system {challenging and computationally expensive}.

\subsection{Order, Disorder, and Correlation Among Chromophores}
We {now} discuss one-body and two-body observables {that can be extracted from the chromophore distributions, in order to further quantify the balance between} order and disorder present in the {chromophore-TMV} system.
There is a simple analogy between our system and one-dimensional classical Heisenberg model of 17 sites with periodic boundary conditions.
In other words, a TMV disk can be reduced down to a lattice with 17 sites and the {orientation of each of the 17 chromophores can be considered analogous to a classical spin on the corresponding site}.
This analogy allows us to utilize one-body and two-body measures that are widely used to quantify order in spin systems.
In passing we note that {the} orientation vectors of TDMs used in the following analyses are all normalized and measured in the monomer frame.
As we analyze only five monomers per sample, {periodic boundary conditions were not applied here}.
\begin{table}[ht]
\centering
\caption{The average of the TDM vector orientation (a.u.) along each cartesian axis in the monomer frame {for the CE and OG systems}.
}
\label{table:onebody}
\begin{tabular}{|c|r|r|r|r|r}
\hline
 & CE-104 & CE-123 & OG-104 & OG-123\\ \hline
$\langle \mu_x\rangle$ & 0.336 & 0.179 & 0.587 & 0.069\\ \hline
$\langle \mu_y\rangle$ & -0.253 & 0.220 & -0.058 & -0.194\\ \hline
$\langle \mu_z\rangle$ & 0.410 & 0.458 & 0.148 & 0.145\\ \hline
\end{tabular}
\end{table}
\begin{table}[ht]
\centering
\caption{The average of the absolute value of the TDM vector (a.u.) along each cartesian axis in the monomer frame {for the CE and OG systems}.
}
\label{table:onebody2}
\begin{tabular}{|c|r|r|r|r|r}
\hline
 & CE-104 & CE-123 & OG-104 & OG-123\\ \hline
$\langle |\mu_x|\rangle$ & 0.558 & 0.389 & 0.824 & 0.159\\ \hline
$\langle |\mu_y|\rangle$ & 0.408 & 0.464 & 0.284 & 0.821\\ \hline
$\langle |\mu_z|\rangle$ & 0.538 & 0.609 & 0.293 & 0.429\\ \hline
\end{tabular}
\end{table}

The one-body {observable} considered here is the average of the magnetization of spins, which in our case is the average of the orientation vector of TDMs, defined as
\begin{equation}
\langle\vec{\mu}\rangle =
\frac{1}{N_\text{samples}N_\text{spins}}\sum_i^{N_\text{samples}}
\sum_\alpha^{N_\text{spins}}
\vec{\mu}_{\alpha}(i).
\label{eq:mu}
\end{equation}
{Here $\langle\vec{\mu}\rangle$ is a normalized TDM vector and the magnitude of each cartesian component of $ \vec{\mu}_{\alpha}(i)$} ranges from 0 to 1.
In the case of ferromagnets, this measure is {sufficient to determine} whether the system is ordered. A small value of $\vec{\mu}$ indicates a disordered phase and a large value indicates an ordered phase. However, in the case of antiferromagnets, a small value of $\vec{\mu}$ is not enough to conclude that it is a disordered phase. This is because a perfect antiferromagnet would exhibit negligible average magnetizations.

Table \ref{table:onebody} shows the {averages of the projections of the TDM orientation vectors} along each cartesian axis in the monomer frame.
CE and OG present a qualitative difference {here, since OG has at least one direction with a very small value.
The small values along the $y$-axis in OG-104 and the $x$-axis in OG-123 are particularly interesting, since} they may indicate an antiferromagnetic ordering along those axes. To further investigate this, we computed $\langle|\vec{\mu}|\rangle$ which is {defined similarly to Eq. \eqref{eq:mu}, with $|\vec{\mu}_{\alpha}(i)|$ replacing $\vec{\mu}_{\alpha}(i)$.
These results} are presented in Table \ref{table:onebody2}.
If there is no difference between $\langle |{\mu}|\rangle$ and $\langle{\mu}\rangle$ then the system is either ferromagnetic or unpolarized, while a significant difference between them suggests an antiferromagnetic ordering. {Both the $y$-component of OG-104 and the $x$-component of OG-123 do show a significant difference, which could therefore be taken as indicating possible antiferromagnetism along those axes.
However, analysis of the two-body correlations below will show that these two components have no ordering.}

We have investigated a two-body correlation {function, (i.e., a two-point correlator), namely 
the spin-spin correlation function, $C_\text{spin}$,}
\begin{equation}
C_\text{spin}
=
\sum_{\langle \alpha \beta \rangle}\langle\vec{\mu}_\alpha\cdot\vec{\mu}_\beta\rangle
=
\frac{1}{N_\text{samples}N_\text{neighbors}}\sum_i^{N_\text{samples}}
\left(
\sum_{\langle\alpha\beta\rangle}
\vec{\mu}^{~i}_{\alpha}
\cdot
\vec{\mu}^{~i}_{\beta}
\right)
.
\label{eq:corr}
\end{equation}
{Here the sum goes over nearest neighbors $\alpha, \beta$. We have included only nearest neighbor correlations in $C_\text{spin}$,} even though the underlying interaction between spins in our case is long-ranged.
This was done on purpose {because the interaction is dominated by nearest neighbor interactions and this truncation leads to a simple interpretation of the physical meaning of $C_\text{spin}$.
In particular, the values of Eq. \eqref{eq:corr} range between -1 and 1, where the limit of 1 corresponds to perfect ferromagnetic order, -1 corresponds to perfect antiferromagnetic order, and 0 indicates either no order, i.e., perfect disorder. We also define $C_\text{spin}^\gamma$ with $\gamma \in \{x,y,z\}$ to quantify these same types of orders along each cartesian axis, with each component defined similarly to Eq. \eqref{eq:corr}, but using TDM components $[\mu_{\gamma}]^{~i}_{\alpha}$}.
\begin{table}[h]
\centering
\caption{Spin-spin correlation functions (Eq. \eqref{eq:corr}) for each system}
\label{my-label}
\begin{tabular}{|c|r|r|r|r|r}
\hline
& CE-104 & CE-123 & OG-104 & OG-123 \\ \hline
$C_\text{spin}$  & 0.433 & 0.390 & 0.386 & 0.035 \\ \hline
$C_\text{spin}^x$ & 0.108 & 0.078 & 0.376 & 0.006\\\hline
$C_\text{spin}^y$ & 0.101 & 0.026 & -0.007 & 0.018\\\hline
$C_\text{spin}^z$ & 0.224 & 0.286 & 0.017 & 0.011\\\hline
\end{tabular}
\label{eq:sstab}
\end{table}

{Table \ref{eq:sstab} shows the values of this two-body observable. We see that CE-104, CE-123 and OG-104 are all considerably more ordered than OG-123. 
Both CE-104 and CE-123 show partial ferromagnetic ordering. This is strongly anisotropic in the case of CE-123 and weakly anisotropic in the case of CE-104, with greater ferromagnetic correlations
along the $z$-axis perpendicular to the plane of the disk in both cases. 
In contrast, Table \ref{table:onebody2} implies that CE-123 has non-negligible orientation along all three $x,y,z$-axes.
Taken together with Table \ref{eq:sstab}, this suggests that CE-123 has weakly correlated partial ferromagnetic order along the $x,y$-axes and more strongly correlated ferromagnetic order that is consequently also greater in extent along the $z$-axis.
Similarly, Tables  \ref{table:onebody2} and \ref{eq:sstab} imply that OG-104 exhibits ferromagnetic ordering along $x$-axis and disorders along the $y,z$-axis.
As illustrated in Figure \ref{fig:tdm_kde}, OG-104 is also strongly confined around the positive $x$-axis.
Therefore, OG-104 is confined and at the same time well-ordered.
OG-104 would thus be a good future candidate for further theoretical studies, since the high degree of both spatial confinement and chromophore ordering means that the entire conformation space need not be explored.
OG-123 is interesting in the sense that Table~\ref{eq:sstab} shows it is disordered along every axis.
However, according to Figure \ref{fig:tdm_kde}, it is nevertheless confined in space.
Although OG-123 is spatially confined by the TMV protein environment, the relative orientation between different chromophores is almost completely random.}


\subsection{Linear Absorption Spectra of Coumarin-TMV double disks}

\begin{figure}[htp!]
    \begin{center}
        \subfigure{
        \includegraphics[width=210pt]{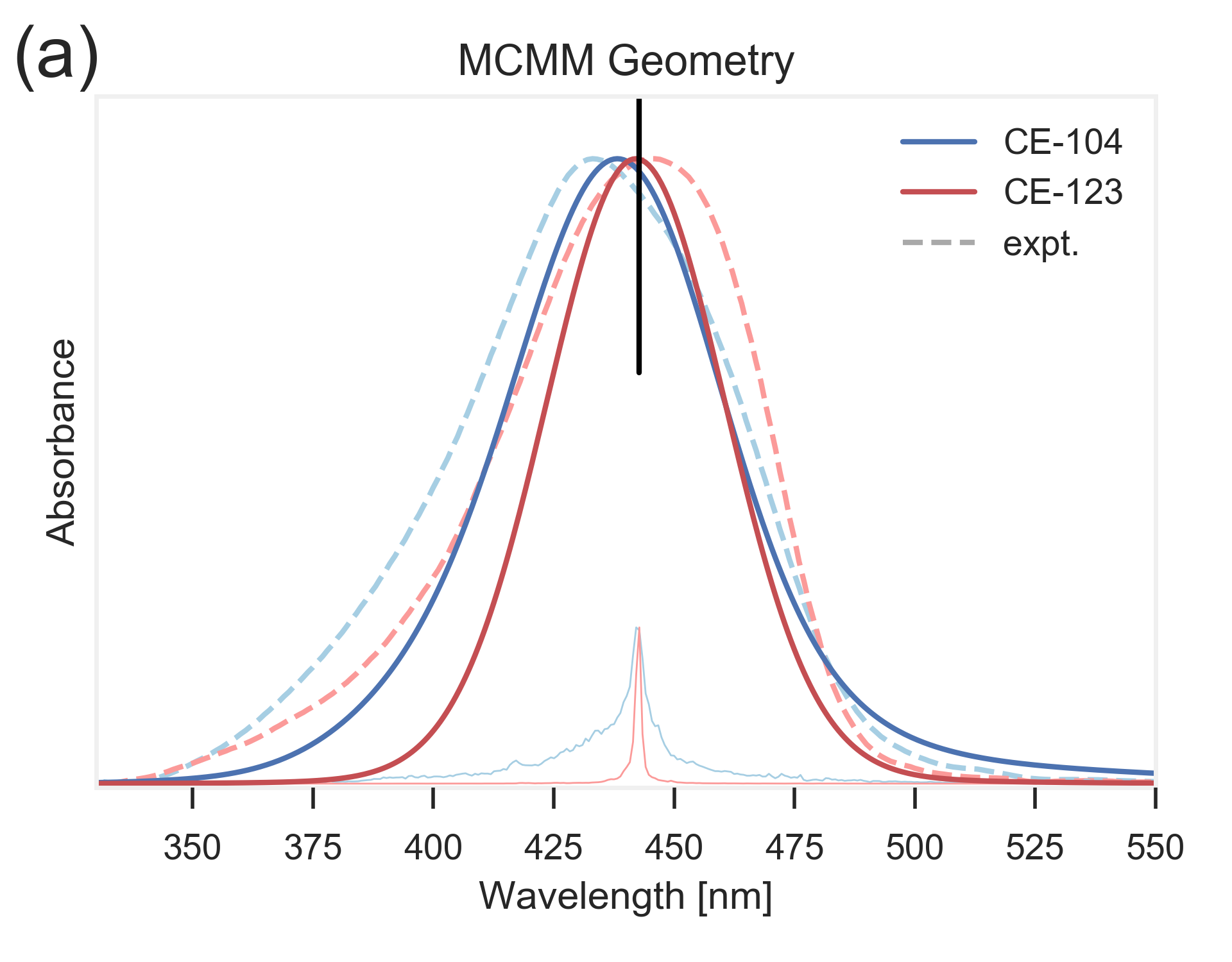}
        }
        \subfigure{
        \includegraphics[width=210pt]{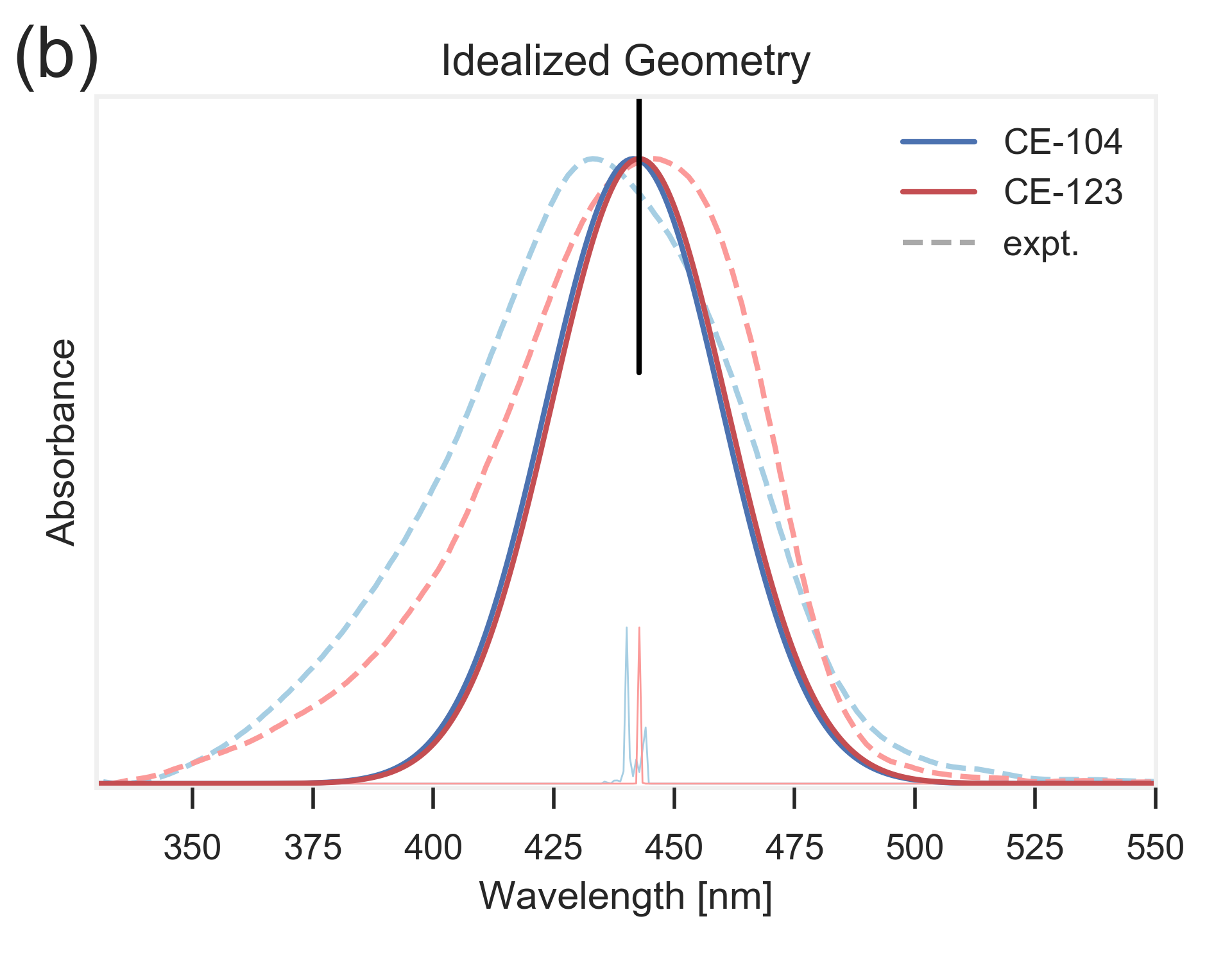}
        }
    \end{center}
    \caption{\small\label{fig:spectra_param}
    Simulations of linear absorption spectra using the TDDFT-improved Hamiltonian parameters in Eq. \ref{eqn:interpolated_coupling}.
    (a) Spectra averaged over the MCMM-geometries (b) spectra of a single idealized, 17-fold symmetric geometry (see Fig. \ref{fig:ideal17ring}).
    Dotted lines are the absorption spectra obtained from experiment, and the vertical black line is the monomer excitation energy $\Delta E^\text{monomer}$.
    The line broadening parameter $\sigma$ is set to 0.12 eV (36.4 nm at a peak position 442.8 nm {corresponding to the shifted monomer energy $\Delta E_1^\text{monomer} = 2.8$ eV (see text)}.
    Below these simulated spectra, we show simulations of the same system using  0.001 eV (0.32 nm at a peak position 442.8 nm).
    }
    \end{figure}

\begin{figure}[htp!]
    \includegraphics[width=350pt]{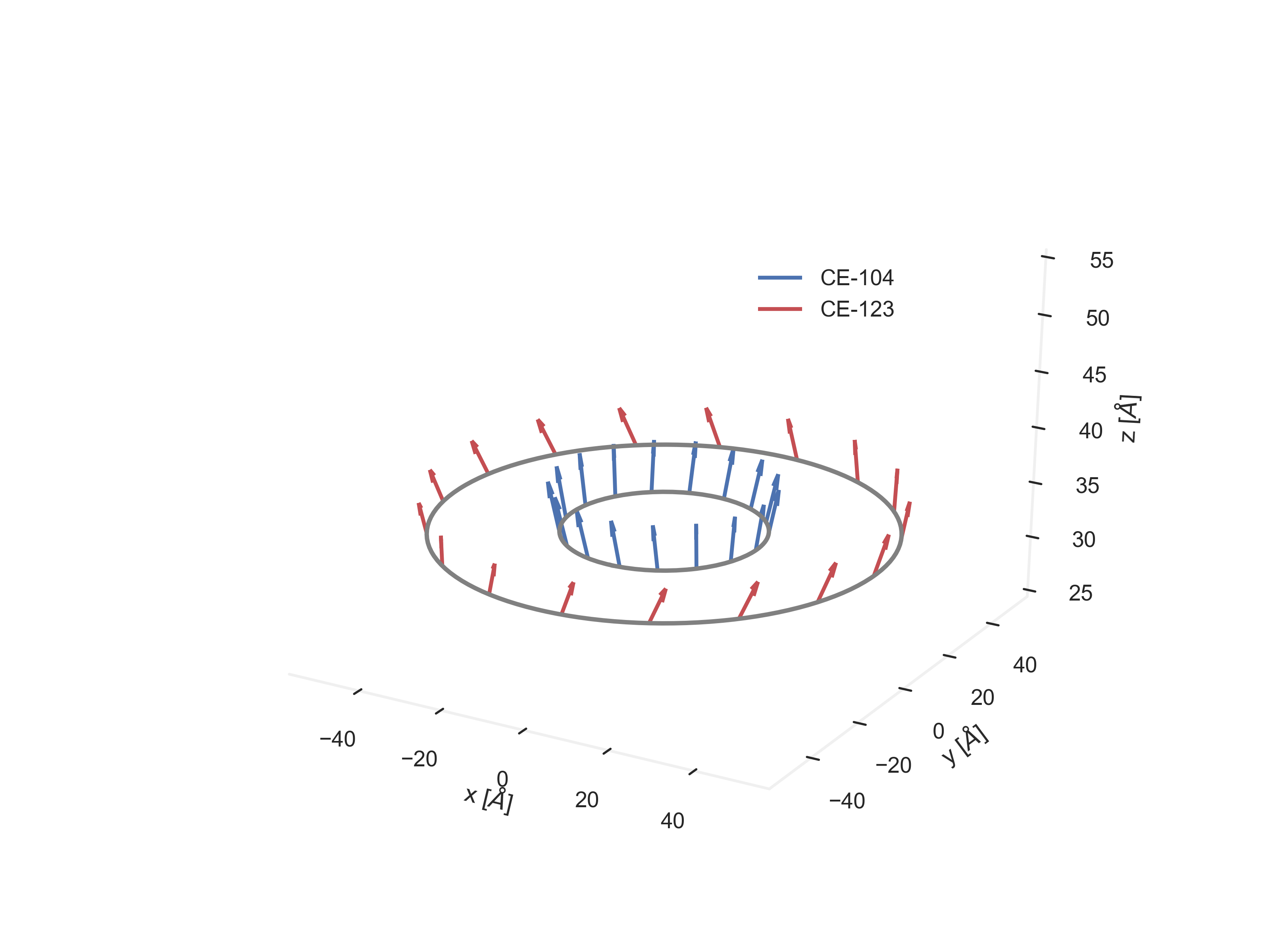}
    \caption{\small\label{fig:geo_sym}
{An} idealized C${}_{17h}$ symmetric geometry where the positions and orientations of the chromophores are obtained from the average over all the MCMM configurations.
    Arrows indicate the orientation of the transition dipole moment. The blue arrows describe the CE-104 system, while the red arrows describe the CE-123 system.
    }
    \label{fig:ideal17ring}
\end{figure}

Figure~\ref{fig:spectra_param} shows simulated linear absorption spectra for the Coumarin TMV-templated aggregates CE-104 and CE-123, and compares these with the corresponding experimental linear absorbance spectra for the fully loaded disks {containing 17 chromophores on each side of the disk}.\cite{Finley2014}  The baseline has been subtracted from the experimental spectra to remove contributions of light scattering.
The left panel, Figure~\ref{fig:spectra_param}(a), shows the simulated spectra resulting from averaging over the MCMM geometries, using the TDDFT-derived Hamiltonian parameters in Eq.~\eqref{eqn:interpolated_coupling} for the CE-104 and CE-123 system as described in Section \ref{ssec:spectra}.  The right panel, Figure~\ref{fig:spectra_param}(b), shows simulated spectra for 
{an idealized 17-fold symmetric structure on a single disk, which has C${}_{17h}$ symmetry (see Fig.~\ref{fig:ideal17ring} and text).}
In all of these Coumarin 343 spectral simulations, we set $\Delta E_1^\text{monomer}=2.8$ eV (442.8 nm), which results in MCMM-averaged {diagonal Hamiltonian energies} $\varepsilon_i = 2.842$ eV and 2.806 eV for CE-104 and CE-123, respectively.
The  $\Delta E_1^\text{monomer}$ energy is considerably smaller than the vacuum TDDFT excitation energy for the CE system (3.47 eV) reported in Table \ref{tbl:coumarin343}, and was shifted from this in order to achieve an optimal average coincidence of the peaks of the simulated spectra with those of the experimental spectra.
This results in a global energy shift {to the diagonal entries of the Hamiltonian} that accounts for the interaction of the Coumarin 343 with the solvent, and the reorganization energy of the chromophores when attached to TMV.
The choice of $\Delta E_1^\text{monomer}$ does not affect our analysis of the difference between the absorption peak locations for CE-104 and CE-123, since our conclusions are all based on relative energy differences.
The line broadening parameter $\sigma$ is set to 0.12 eV (36.4 nm at {a peak position 2.8 eV $\equiv$ 442.8 nm})  for all simulated spectra.  This parameter provides a qualitative fit to the overall spectral distribution and represents implicit dependency on \revision{both} the homogeneous line broadening {that results} from coupling to vibrational degrees of freedom \revision{and the inhomogeneous line broadening arising from static disorder in the transition energies}.

We note that the {linewidth parameter $\sim 36$ nm} 
is the same order of magnitude as that of typical photosynthetic chromophores in solution~\cite{Blankenship2014} and just as for aggregates of such chromophores in light harvesting, it is larger than the inter-chromophore coupling $J$ (see Table \ref{tbl2:ham_avg}).
At the bottom of each of these plots, we also show the same spectral simulations with $\sigma$ set to 0.001 eV (0.32 nm at a peak position 442.8 nm).
Since this broadening value is lower than the resolution of the x-axis, these insert plots can be interpreted as histograms of the stick spectra from ground to discrete
excited state energies,
which corresponds to heterogeneous line broadening.


We first discuss the simulated spectra resulting from averaging over the MCMM geometries, panel (a), and then the spectra resulting from averaging over instantaneous idealized C${}_{17h}$ symmetric geometries,  panel (b).

Figure~\ref{fig:spectra_param}(a) shows that the distribution of the lines in the stick spectrum for the CE-123 system is considerably narrower than that for the CE-104 system.
This is due to the fact that in all geometric configurations of the CE-123 system the chromophores are further apart than in the CE-104 system, resulting in smaller off-diagonal couplings and a consequent smaller range of the electronic excited state eigenvalues.
In the limit of infinite separation, these excited state eigenvalues become degenerate and would yield a delta function spectrum.
In comparison, the CE-104 system (blue line) has a much wider stick spectrum, showing non-zero absorption between 400-475 nm.
{This is due in part to the greater range of the electronic excited state eigenvalues for CE-104 ($\sim$ 2.22 eV), and also to the greater contribution of static disorder in the chromophore positions and orientations (the range of individual excited state eigenvalues for CE-104 averaged over all geometry instances is 0.80 eV). Both of these factors contribute to the significant inhomogeneous broadening of the CE-104 spectral absorption.
It is also apparent that the stick spectrum distribution of CE-104 shows a signifiant asymmetrical weight towards higher energies, while that of CE-123 is more symmetrical.  This results in a greater asymmetry in the convolved spectrum for CE-104, consistent with the greater broadening of the experimental spectrum on the blue side of the peak. This greater asymmetry for CE-104 also contributes to the overall blue shift of the CE-104 relative to the CE-123 system.}

{To obtain more insight into the effect of the chromophore geometries on the absorption line shape, Figure~\ref{fig:spectra_param}(b) {shows a computed spectrum} for an ideal C${}_{17h}$ symmetric geometry that is constructed by setting  the positions and orientations of the chromophores to  the average values over all the MCMM configurations. Here the spread in the stick spectrum now derives solely from the excitonic splitting of the 17 excited state energies, since all chromophores have the same position and orientation on the TMV disk.
{The resulting idealized geometry on the top half of a double disk} is shown for both the CE-104 and CE-123 systems in Figure~\ref{fig:geo_sym}.
The symmetry of these geometries implies that every nearest-neighbor pair of chromophores has an identical geometry relative to each other, so no pair couples more strongly than any of the others.
For such idealized geometries, the resulting Hamiltonian will yield eigenvalues with a smaller spread than the range of values obtained from sampling with MCMM. Consequently, we see a tighter spectrum for CE-104 in Figure~\ref{fig:spectra_param}(b) than in Figure~\ref{fig:spectra_param}(a).}

\begin{table}[]
\centering
\begin{tabular}{|c|r|r|r|r|r|r|r}
\hline
          & $\lambda_\text{CE-104}^\text{avg}$ [nm] & $\lambda_\text{CE-123}^\text{avg}$ [nm] & $\Delta_\text{avg}$  [nm] & $\lambda_\text{CE-104}^\text{max}$ [nm] & $\lambda_\text{CE-123}^\text{max}$ [nm] & $\Delta_\text{max}$  [nm] \\ \hline
Experiment~\cite{Finley2014}      & 429.28  &  435.14  & 5.86  & 433.07  & 445.07  & 12.00   \\ \hline
MCMM  & 438.32  & 442.78  & 4.46   & 437.46  & 442.03  & 4.57  \\ \hline
Idealized & 442.42 & 443.60  & 1.18  & 445.13  & 443.06  & -2.06\\ \hline
\end{tabular}
\caption{Mean and Maximum Wavelength of Absorbance. $\Delta_\text{avg}$ is the difference between CE-104 and CE-123 mean wavelengths ($\lambda_\text{CE-123}^\text{avg} - \lambda_\text{CE-104}^\text{avg}$), and $\Delta_\text{max}$ is the difference between CE-104 and CE-123 peaks ($\lambda_\text{CE-123}^\text{max} - \lambda_\text{CE-104}^\text{max}$).}
\label{tbl:absorbance}
\end{table}

%

We quantify the differences between the CE-104 and CE-123 spectra in Figure~\ref{fig:spectra_param} by calculating the
{peak and mean wavelength of absorption for each spectrum, where the latter is give by} the weighted average:
\begin{equation}
    \lambda^\text{avg} = \frac{\sum_\lambda \text{Abs}(\lambda) * \lambda }{ \sum_\lambda \text{Abs}(\lambda)}.
\end{equation}
The values are shown in Table~\ref{tbl:absorbance} for {both the MCMM averaged spectra and the idealized spectrum}, where they are also compared with the
peak absorption values obtained from experimental spectra.\cite{Finley2014} The peaks in the experimental spectrum exhibit a blue shift when going from the CE-123 system to the CE-104 system.  Taken together with the narrow distribution of the individual excitonic absorptions in the CE-123, this is consistent with the lack of any significant interaction-induced shift of the chromophores in the CE-123 system, while the more closely packed ring of aggregates in the CE-104 system behaves as an H-aggregate.
Table~\ref{tbl:absorbance} shows that the simulation using the MCMM-derived geometries is better able to reproduce this relative blue shift of the experimental spectra than the simulation using the idealized 17-fold structure.
While both the simulations using the MCMM geometries and the idealized geometries exhibit a blue shift, the simulation that incorporates disorder is able to match this trend much better.
We conclude that the disorder in the geometry of the CE-104 and CE-123 systems is an important feature of these systems, and it is important to account for this disorder in order to model {the optical properties of TMV-templated chromophore aggregates} accurately.

\section{Conclusions}
\label{sec:conclusions}

In this work, we present a protocol that generates conformations using a Monte Carlo multiple minima (MCMM) conformation search algorithm, parametrizes a semiempirical tight-binding Hamiltonian based on ab-initio TDDFT calculations, and combines {these} to generate a linear absorption spectrum that can be directly compared to experiments.

{We} applied this protocol to study a recently synthesized artificial light harvesting system consisting of chromophores attached to a tobacco mosaic virus (TMV) protein.
{We studied Coumarin 343 together with a linker, and Oregon Green 488, both of which were} attached to the 104 and 123 sites on the TMV protein.
The resulting four systems, CE-104, CE-123, OG-104, and OG-123, were studied with MCMM and we obtained a wide array of local minima.
We characterized those conformers using the orientation of the transition dipole moment and center-of-mass of dyes attached to the TMV protein.
Such a characterization led to {a qualitative and quantitative} understanding of structural order and disorder associated with the dyes.
CE-104 and CE-123 both exhibit a very broad geometric distribution, which makes any {more detailed theoretical study relatively} intractable.
OG-104 and OG-123 are relatively spatially well confined, but OG-123 is more disordered than is OG-104 in terms of the spin-spin correlation function discussed in the main text.
For future studies, we {therefore conclude that OG-104 will likely be the system most suited for more detailed theoretical study}.

Lastly, we {combined the wide array of conformations found through MCMM with a semiempirical tight-binding Hamiltonian for the Coumarin-linker system to calculate linear absorption spectra of CE-104 and CE-123, and compared this with experimental spectra}.
We observed that it is necessary to account for the proper {distribution over geometries of conformations to properly reproduce the experimentally observed blue shift of the CE-104 system relative to the CE-123 system. We also confirmed the greater asymmetry of the lineshape of the CE-104, with detailed analysis showing that this derives largely from the greater distribution of the excitonic energies for CE-104, due to the closer distance between chromophores}.
{It is encouraging that a qualitatively accurate spectrum could be obtained in-silico from MCMM calculations together with a tight-binding Hamiltonian.  A challenge for further work is to incorporate the interaction with vibrational degrees of freedom, as well as to develop more reliable and economical ways to generate energy minima and thereby increase the conformational sampling.}.



\section{Acknowledgement}
This work was supported by the DARPA QuBE (Quantum Effects in Biological Environments) program under contract number N66001-10-1-4068. The views expressed are those of the authors and do not reflect the official policy or position of the Department of Defense or the US Government.
J. L. thanks POSTECH for the exchange student scholarship that made early stages of this work possible.

\section{Table of Contents Graphic}
\begin{figure}[h!]
\includegraphics[scale=0.35]{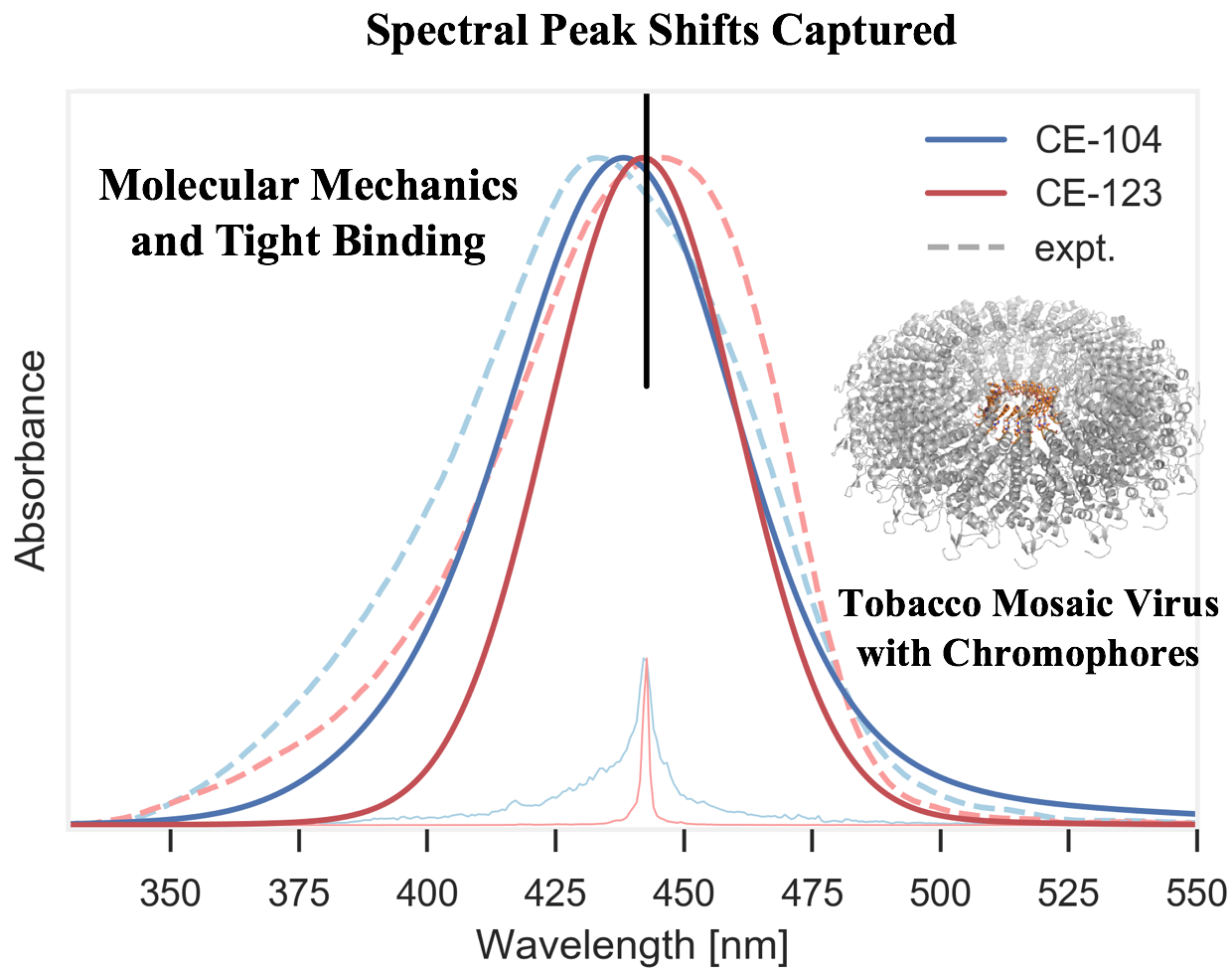}
\end{figure}

\bibliography{mol_mec_revision}
\bibliographystyle{achemso}
\end{document}